**This is the accepted manuscript (postprint) of the following article:**
S. Rahimipour, B. Rafiei, E. Salahinejad, *Organosilane-functionalized hydrothermal-derived coatings on titanium alloys for hydrophobization and corrosion protection*, Progress in Organic Coatings, 142 (2020) 105594.
https://doi.org/10.1016/j.porgcoat.2020.105594

# Organosilane-functionalized hydrothermal-derived coatings on titanium alloys for hydrophobization and corrosion protection

S. Rahimipour, B. Rafiei, E. Salahinejad*

Faculty of Materials Science and Engineering, K. N. Toosi University of Technology, Tehran, Iran

## Abstract

This work focuses on the structure, wettability and corrosion behaviors of Ti-6Al-4V alloy after roughening treatments in different concentrations of NaOH aqueous solutions followed by low surface energy hexadecyltrimethoxysilane (HDTMS) coating. In this regard, scanning electron microscopy, contact angle measurements, potentiodynamic polarization and electrochemical impedance spectroscopy were used to characterize the samples. In contrast to hydrophilicity caused by the hydrothermal alkaline treatments, the subsequent HDTMS coating donated considerable hydrophobicity. Typically, the highest sessile water contact angle (about 147°) was obtained for the sample treated in 3 molar NaOH solution followed by the HDTMS coating. In addition, the alkaline treatment reduced the corrosion resistance of the surface in a NaCl aqueous solution; however, the HDTMS hydrophobization process improved it significantly. It is eventually concluded that the coupled use of alkaline treatment and HDTMS functionalization can be further considered for moisture-exposed applications of Ti-based alloys.

* Corresponding Author: Email Address: <salahinejad@kntu.ac.ir>

## 1. Introduction

Titanium-based alloys have various applications especially in aerospace, automobile, marine and medical industries, mainly due to their high specific strength and corrosion resistance in most working conditions. The high corrosion resistance of these alloys is attributed to the formation of a spontaneous oxide passive layer on their surface when exposed to the service environments [1]. On the one hand, this surface layer is more or less susceptible to high concentrations of chloride, leading to localized corrosion attack. Thus, there is still a need to improve the corrosion resistance of these alloys, particularly in chloride-containing environments like seawater and body for marine and medical applications, respectively [2, 3]. On the other hand, some applications of these alloys suffer from the adsorption and subsequent freezing of environmental moisture, leading to the loss of their relevant functions.

In general, corrosion and freezing can be avoided if water from the environment (moisture) cannot join together to create an electrolytic film on the surface [4, 5]. This feature can be achieved by making the surface superhydrophobic and hydrophobic, giving rise to water repellency, anti-corrosion, self-cleaning and anti-mass properties [6, 7]. Essentially, the surface wettability is controlled by surface free energy and roughness, where the increase of roughness along with the reduction of surface energy can increase hydrophobicity [8]. Concerning the corrosion protection of titanium alloys through hydrophobization, the successful deposition of pulsed-laser hopeite [9], plasma-sprayed hydroxyapatite [10], sputtered tantalum nitride [11] and perfluorooctyltriethoxysilane after sandblasting and anodic oxidation [12] are noticeable.

Hexadecyltrimethoxysilane (HDTMS) is a fluorine-free organosilane which offers hydrophobization if deposited on a desirably rough surface. Regarding HDTMS functionalization for the corrosion protection of metallic materials, reports on Cu-Zn [13], Mg-Al-Zn [14], Ni-Ti [15] and Al-Mg [16] alloys are noticeable. This work, for the first time, aims to use this low surface free energy silane coupled to a simple, inexpensive and flexible method of roughening on Ti alloys. In this regard, NaOH hydrothermal treatments are employed prior to HDTMS coating. The alkaline treatment forms a porous sodium-containing layer on the surface via placing the metal in an alkaline solution for a specific period. Typical factors affecting the final characteristics are the alkaline solution concentration, immersion temperature, immersion time and subsequent thermal treatment [17, 18]. The application of alkaline-treated surfaces with a rough and oxide nature as the substrate of HDTMS can also ensure the efficiency of this silane towards hydrophilicity, since the direct coating of HDTMS on metallic surfaces results in a weak hydrophobic efficiency [19, 20].

## 2. Materials and methods

### *2.1. Materials*

The substrates were commercially available Ti-6Al-4V alloy disks of 10 mm in diameter and of 2 mm in thickness. In order to achieve a variety of roughness, alkaline treatments in aqueous solutions of sodium hydroxide (NaOH, Merck, Germany, 99.0%) and to reduce the surface free energy, hexadecyltrimethoxysilane (HDTMS, $H_3C(CH_2)_{15}Si(OCH_3)_3$, Sigma-Aldrich, ≥85%) were used. Distilled water obtained from an AQUA ARA machine, ethanol (Ethyl alcohol, Nasr Alcohol, 96%) and acetone (Sasol) were also used as the solvents of the surface engineering solutions and for washing. Acetic acid (Mojallali industrial chemical complex, ≥99%) was also utilized to reduce the pH of the HDTMS coating solution.

### *2.2. Initial surface preparation*

The disks were first polished by sandpapers of 110, 220, 360, 600, 800 and 1000 # to obtain a smooth and oxide-free surfaces. Then, they were cleaned by acetone, ethanol and distilled water for 10 min in an ultrasonic bath (Parsonic 7500s).

### *2.3. Alkaline treatment*

To obtain a range of roughness, according to Refs. [17, 18], the samples were immersed in 100 ml of the alkaline solutions with the NaOH concentrations of 1, 3 and 5 molars at 60 ° C for 24 h. Afterwards, the samples were washed in ethanol and distilled water and dried at 60 ° C for 24 h to avoid cracking [21].

### *2.4. HDTMS coating*

Based on Refs. [15, 20], 3% of HDTMS was added to a solution of 90 ml of ethanol and 10 ml of distilled water. The solution was stirred for 60 min by a magnetic stirrer after adjusting pH at 5 with acetic acid. The samples were immersed in this solution for 2 h, followed by rinsing with ethanol and distilled water. After drying at ambient temperature, they were cured at 120 ° C for 60 min. Figure 1 depicts the schematic of the surface treatments and Table 1 tabulates the samples codes.

### *2.5. Structural evaluations*

The surface morphology of the samples was studied by a field emission scanning electron microscope (FESEM, MIRA3TESCAN-XMU) at an acceleration voltage of 10 kV after gold sputtering. Also, the energy-dispersive X-ray spectroscopy (EDS) analysis was used to determine the chemical composition of the surfaces.

### *2.6. Water contact angle measurements*

5 µl of distilled water was dropped on the samples with three repetitions and photographed by a professional digital camera from side views. The sessile water contact angles were measured by tangent lines to the interfaces at the three-phase points and averages were assigned to each sample.

### *2.7. Electrochemical corrosion experiments*

In order to study the effect of the alkaline treatment and hydrophobization on the corrosion behavior of the surfaces, a potentiostat/galvanostat device (Vertex, Ivium Technologies), platinum counter electrode, Ag/AgCl reference electrode and surface area of 0.2 $cm^2$ were utilized. To do so, the samples were first soaked in a 3.5 wt% NaCl solution for 60 min, providing stable open circuit potentials (*OCP*s). Electrochemical impedance spectroscopy (EIS) experiments were conducted over a frequency range of 10 kHz to 10 MHz with the potential amplitude of 10 mV at the *OCP*. Potentiodynamic polarization tests were also performed at a scan rate of 0.1 mV/s and a range of -1 V to +1 V. The electrochemical data were analyzed by the IviumSoft software, where the corrosion current density of the samples was obtained by extrapolating the Tafel tangent and obtaining the intersection point of cathodic and anodic curves.

## 3. Results and discussion

### *3.1. Structural characterization*

Figure 2 represents the FESEM micrograph of the non- and alkaline-treated samples. According to the micrographs, a considerable dependency of the surface morphology on the NaOH concentration of the solution is evident. The surface of Sample 3 is covered by a porous quasi-network, where each wall is almost 15 nm in thickness and each grid is about 50-100 nm in hollow size. By increasing the concentration of NaOH to 3 molar (Sample 5), the walls and grids of the networks are ripened and reach about 50 nm and 400-700 nm in size, respectively. When the concentration of the alkaline solution reaches 5 molar (Sample 7), some rods of 50-180 nm in diameter and of 500-1000 nm in length are protruded from the networks. These morphological dependency on the NaOH concentration is as a result of variations in the nucleation and growth kinetics of the formed surface scales [22, 23].

The chemical composition of the untreated and alkaline-treated surfaces was also analyzed by EDS (Figure 3 and Table 2). As can be seen, by increasing the concentration of the alkaline treatment solution, the amounts of titanium, aluminum and vanadium, as the main elements of the titanium alloy substrate, are reduced. This can be explained by the morphological growth of the surface porous networks with increasing the NaOH concentration, as described in Figure 2. It should be, on the one hand, noted that the decrease in the level of titanium is lower than the two other elements, which is an evidence for the presence of titanium in the surface layer formed during the alkaline treatments. On the other hand, the surface contents of oxygen and sodium are enhanced with the NaOH concentration, suggesting that the porous networks are mainly composed of Na, O and Ti. This is in agreement with X-ray diffraction analyses indicating that alkaline-treated surfaces consist of sodium titanate [24, 25].

*3.2. Water contact angle analyses*

Figures 4 and 5 represent the results of the water contact angle measurements on the samples. On the one hand, comparing the wettability of Sample 1 with Samples 3, 5 and 7, it is realized that the alkaline treatment increases the hydrophilicity of the Ti alloy surface. This is due to the evolution of both surface morphology and chemical composition accompanied by the formation of a hydrated gel layer of sodium titanate with surface Ti-OH groups [26, 27]. The decrease in the water contact angle after the alkaline treatments is also attributed to the surface morphology evolution (Figure 2). This is in good agreement with the Venzel's model [28] stating that by increasing roughness, hydrophilic surfaces will become more hydrophilic and hydrophobic surfaces will become more hydrophobic. Other studies have also confirmed that the increase in the roughness of titanium-based alloys enhances the level of wettability [29, 30]. The higher hydrophilicity of Sample 5 than Sample 7 suggests the optimal surface morphology of the former sample.

On the other hand, it is observed that the HDTMS coating offers considerable hydrophobicity owing to the reduction of the surface energy. The comparison of the HDTMS-coated samples also infers that the alkaline-treated substrates yield more hydrophobicity with respect to the untreated substrate, due to the increase in roughness explained by the Venzel's model above. It should be noted that due to the molecular thickness of the HDTMS coating, no changes in roughness occur after this coating process [20]. The higher hydrophobicity of Sample 6 than Sample 8 is an evidence of the optimal surface morphology of Sample 6, which is in good agreement with the higher wettability of Sample 5 than Sample 7. Indeed, the water droplet does not easily penetrate into the surface cavities of Sample 6 and some air remains inside the cavities. Nonetheless, the larger size of holes on the surface of Sample 8 can change the wetting behavior from the Cassie-Baxter to Wenzel states. In this case, the air below the droplet is thermodynamically unstable [31, 32], the droplet penetrates into the cavities and the sessile water contact angle is reduced. In addition, it should be noted that wettability is enhanced when the height-to-width ratio of ups and downs increases [33], where the capillary effect helps to reduce the contact angle of water at high levels of roughness [34] like Sample 8.

According to the literature [19, 20], the direct coating of HDTMS on metallic surfaces does not give rise to considerable hydrophobicity, due to a weak linkage of the coating and substrate. When HDTMS is deposited on the alkaline-treated surfaces, surface hydroxyl groups of the substrate attack silicon atoms of silane, forming Si-OH bonds and causing the release of $OCH_3$ groups. Then, another anion of the additional hydroxyls formed in the previous reaction takes proton from the silane groups in water. Further, oxygen atoms of silane, with a negative charge, react with the surface atoms of the samples, leading to the adhesion of HDTMS to the surface [35, 36]. The reduction of surface energy and thereby hydrophobicity is also attributed to the presence of non-polar C16 alkyl chains on the surface [30], which ultimately leads to the repulsion of the non-polar molecules of the carbon chain and the polar molecules of water [37].

### *3.3. Corrosion studies*

Figure 6 indicates the potentiodynamic polarization curves of the samples in terms of potential ($E$) vs. current density ($i$). It is observed that the hydrophobization process by the HDTMS coating shifts the diagrams to the left (less current densities), which is indicative of improved corrosion resistance. This is due to the electrolyte repulsion, actual electrolyte/surface contact area reduction and corrosive ions infiltration inhibition, because an air layer is trapped between the electrolyte and hydrophobized surface. Table 3 also lists quantitative data extracted from the polarization diagrams. The protection efficiency of the coatings ($\eta$) was also calculated by Equation 1 [38, 39] and listed in Table 3.

$$\eta = (1 - \frac{i_{corr}}{i^0_{corr}}) \times 100 \qquad (1)$$

where $i^0_{corr}$ is the current density of the untreated sample (Sample 1).

The comparison of the corrosion potential of the samples, as a criterion of corrosion tendency from the viewpoint of thermodynamics, does not exhibit a meaningful trend, which is attributed to the different nature of the surfaces showing different roughness and wettability [40]. Therefore, for the more logical comparison of the electrochemical properties of the samples, the corrosion current density is used. It can be seen that Samples 3, 5 and 7, i.e. those subjected to the alkaline surface treatments, present higher current densities than Sample 1, i.e. the untreated sample. This is compatible with other studies that conclude alkaline treatments reduce the corrosion resistance of Ti-based alloys, because the gel layer formed on the alkaline-treated samples exhibits less corrosion resistance than the passive layer of titanium oxide [41]. Concerning the HDTMS-coated samples, it can be seen that Samples 4, 6 and 8, which have been subjected to the alkaline treatments prior to coating, present lower current densities than Sample 2 which has been merely polished prior to coating, due to the higher hydrophobicity (Figures 4 and 5). Also, from comparing the electrochemical results of Sample 4 with Samples 6 and 8, the increase in the corrosion resistance by increasing the concentration of the NaOH solution.is attributed to the increase of hydrophobicity. Other studies have indicated that the corrosion protection of hydrophobic materials is due to trapped air packets at solid/liquid contact surfaces, which prevents the infiltration of corrosive ions to the substrate surface [42, 43].

The Nyquist and bode plots of the samples are depicted in Figure 7. According to Figure 7a, the feature of the Nyquist plots can be divided into two groups. The plots for Samples 1, 2, 3, 5 and 7 exhibit semicircular arcs which interrupts the horizontal axis of the real impedance. The plots for the other samples, i.e. those subjected to the alkaline treatment coupled with HDTMS coating, present a semicircular arc which does not interrupt the horizontal axis in the scanned frequency range. In the first category, it can be seen that the alkaline-treated samples exhibit lower arc radii than the non-treated sample. In the second group, by increasing the concentration of the alkaline solution, the radii are enhanced, where the radius for Sample 8 tends to infinity. Furthermore, the Bode impedance values ($Z$) of the first group samples are lower those of the second group samples (Figure 7b). Considering the fact that the semicircular radius of Nyquist plots and the impedance value of bode plots at low frequencies are proportional to corrosion resistance, it is concluded that the Nyquist and Bode analyses qualitatively confirms the corrosion resistance ranking realized by the polarization analysis.

According to Figure 7c, the Bode phase angle amount of the samples first increases with frequency, reaches highest values at medium frequencies and then decreases to near-zero angles. Such behavior is a sign of a single time-constant behavior of EIS, which is compatible with the single semicircular Nyquist plots (Figure 7a). This points out that none of the samples are exposed to the breakdown of the protective surface layer and thereby serious pitting corrosion during the EIS experiments, where inductive loops are generally detected at medium frequencies for surfaces suffering from substantial corrosion deterioration [13, 14]. The near-zero part of the Bode phase angle plots at high frequencies, which corresponds to the constant value of impedance (Figure 7b), also displays the electrolyte resistance. At lower frequencies, by decreasing the frequency, the logarithm of impedance with a nearly fixed gradient increases, which is related to the maximum parts of the Bode phase angle plots. This is also a characteristic of the capacitive behavior of double/protective layers. It suggests the equivalent electrical circuit of Figure 7d, in which the capacitor was replaced by a constant phase element (CPE) because the maximum of the Bode phase angle curve deviates from the ideal value (90°). The same equivalent electrical circuit has been used for EIS modeling of Ti [44, 45], alkaline-treated Ti [46] and other hydrophobized surfaces [47, 48]. The impedance of a CPE is calculated by the following equation [40, 49-51]:

$$Z_{CPE} = \frac{1}{Y_0(j\,\omega)^n} \qquad (2)$$

where $Y_0$ and $n$ are the independents of frequency, $\omega$ is angular frequency ($\omega = 2\pi f$, where $\pi = 3.14$ and $f$ is frequency) and $j$ is $\sqrt{-1}$. Also, in this circuit, $R_s$ is related to the electrolyte resistance due to the potential loss between the jaw tip and electrode surface. Also, $R_{ct}$ is the electrical resistance associated with the transfer of electric charge from the electrolyte into the films existing on the samples surfaces. Table 4 tabulates the fitting parameters of the EIS results with the proposed electrical circuit. The proximity of the calculated $n$ values to unit suggests that the CPEs behave similar to capacitors. Thus, the equivalent capacitance values of the double layers ($C$) were also calculated by Equation 3 [40] and listed in Table 4.

$$C = (Y_0 \cdot R_{ct}^{1-n})^{\frac{1}{n}} \qquad (3)$$

To understand the mechanisms which control the corrosion behavior of the different processed surfaces, the samples can be categorized into three groups:

i) Sample 1: the surface of this sample is instantly covered by a passive oxide film when exposed to the electrolyte [41]. This film acts as a resistive barrier against both the physical access of corrosive ions to the metallic surface and the electron transfer due to the semiconductive nature of the passive film.

ii) Samples 3, 5 and 7: on the one hand, the porous nature of the alkaline-treated surfaces enhances the actual contact area of the electrolyte and surface. On the other hand, these sodium titanate surfaces are hydrated and transformed to gel layers when exposed to the electrolyte [26, 27]. Although these gel layers moderately protect the metallic surface against corrosion, the protection efficiency of these layers with high surface areas is lower than the passive oxide layer.

iii) Samples 2, 4, 6 and 8: the hydrophobization processing of the surfaces provides an air layer between the liquid electrolyte and solid surface when exposed to the electrolyte [42, 43]. This leads to the decrease of the actual contact area of the electrolyte and surface. The improved corrosion resistance of these samples is due to both the insulating character of the air layer and the reduced contact area. These mechanisms are further activated by increasing the concentration of the alkaline solution and thereby hydrophobization.

According to Equation 4, the capacity is determined by the thickness of the protective layers ($d$), the actual contact area ($A$) and/or the dielectric constant of the protective layers ($\varepsilon$), where the latter depends on porosity, wettability and chemical composition of the protective layers.

$$C = \frac{A\varepsilon\varepsilon_0}{d} \quad (4)$$

where $\varepsilon_0$ is the dielectric constant of vacuum [52, 53]. Thus, the compromise of these parameters for the passive, gel and air layers controls the $C$ values for the samples. Since higher $R_{ct}$ and lower $C$ are indicative of the improved corrosion resistance, it is realized form Table 4 that the alkaline treatment and HDTMS processing reduces and increases the corrosion resistance, respectively. It is also noticeable that these decreasing and increasing variations are intensified by increasing the concentration of the alkaline solution. This suggests that the EIS-drawn conclusions are in good agreement with the potentiodynamic polarization results.

## 4. Conclusions

In this research, the effect of alkaline treatment and HDTMS coating on the structure, wettability and corrosion behavior of Ti-6Al-4V alloy was investigated. The surface morphology of the alkaline-treated samples included nano- and submicron-scales holes and rods, with an increase in the size of the cavities with the alkaline solution concentration. The hydrothermally-treated samples after HDTMS coating showed hydrophobic behaviors, where the highest wettability angle was obtained for the sample alkaline-treated in a 3 molar NaOH solution followed by the HDTMS coating (equals to 147.1 ± 1.4°). The alkaline treatment reduced, but the HDTMS coating improved the corrosion resistance in the chloride-containing electrolyte. It is finally concluded that the couple of alkaline treatment and HDTMS coating, as a straightforward and flexible approach, can be further considered for moisture-exposed applications of titanium alloys.

## Figures

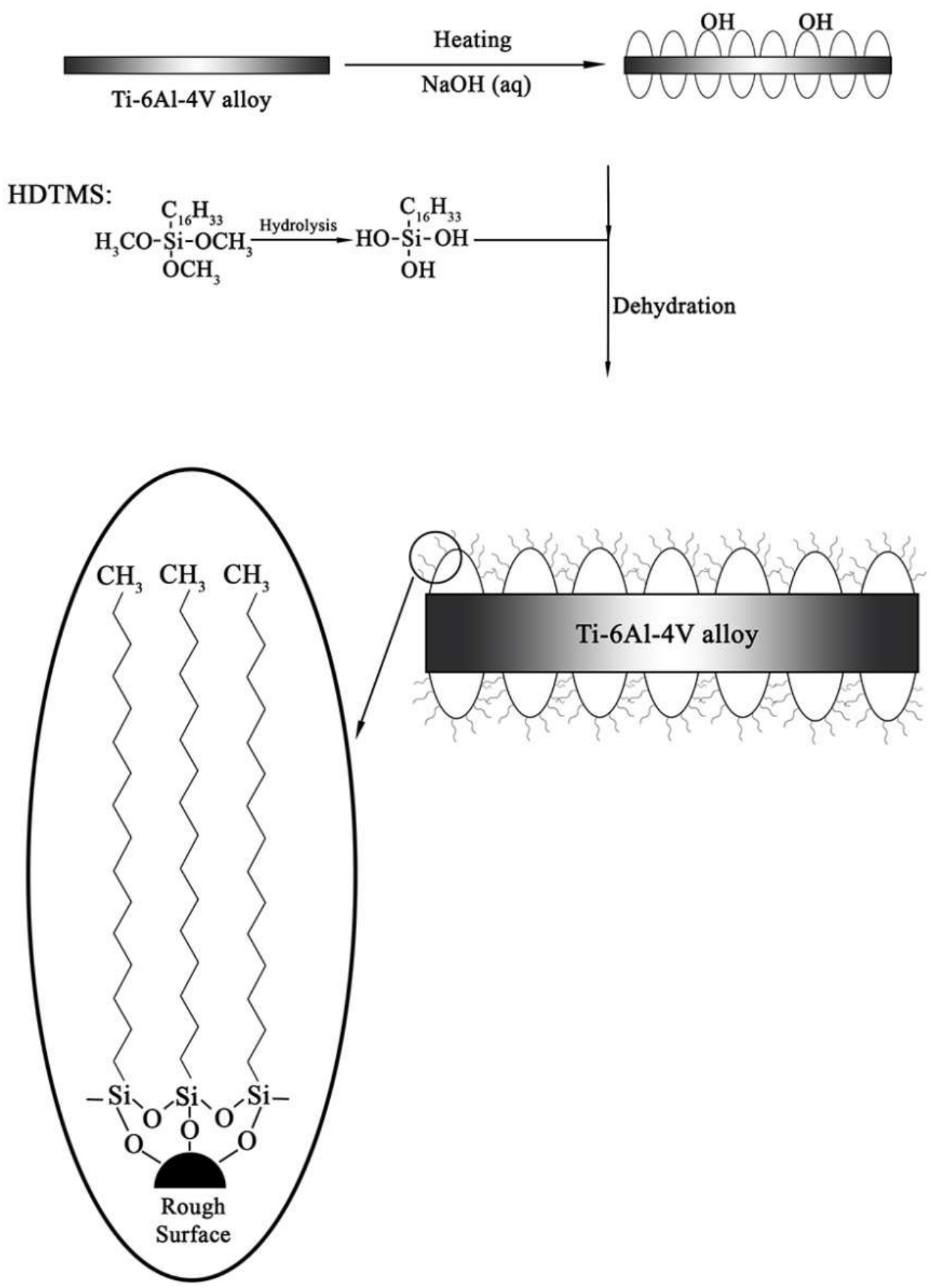

Figure 1. Schematic illustration of the hydrophobization process used.

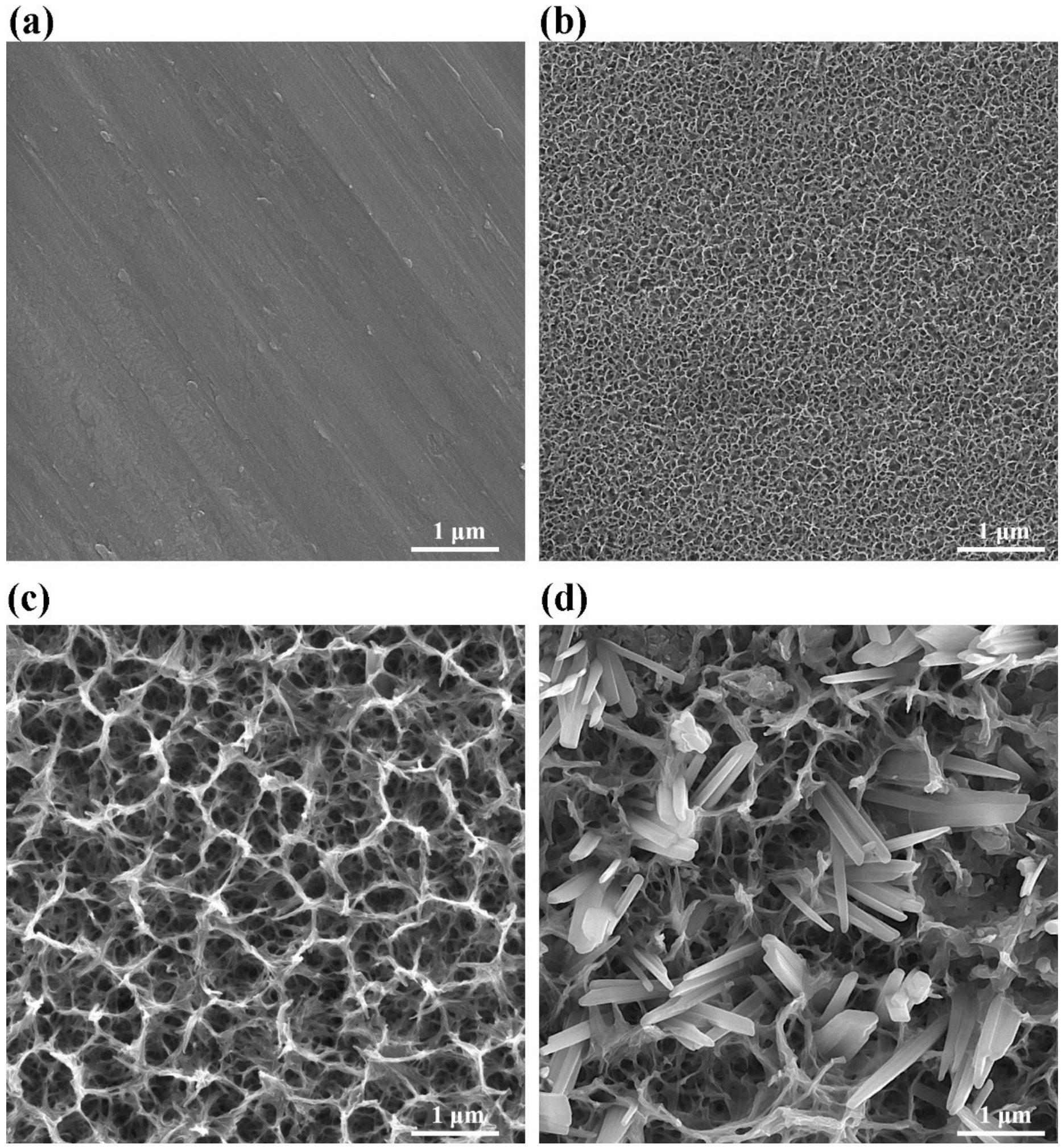

Figure 2. FESEM micrographs of Samples 1 (a), 3 (b), 5 (c) and 7 (d).

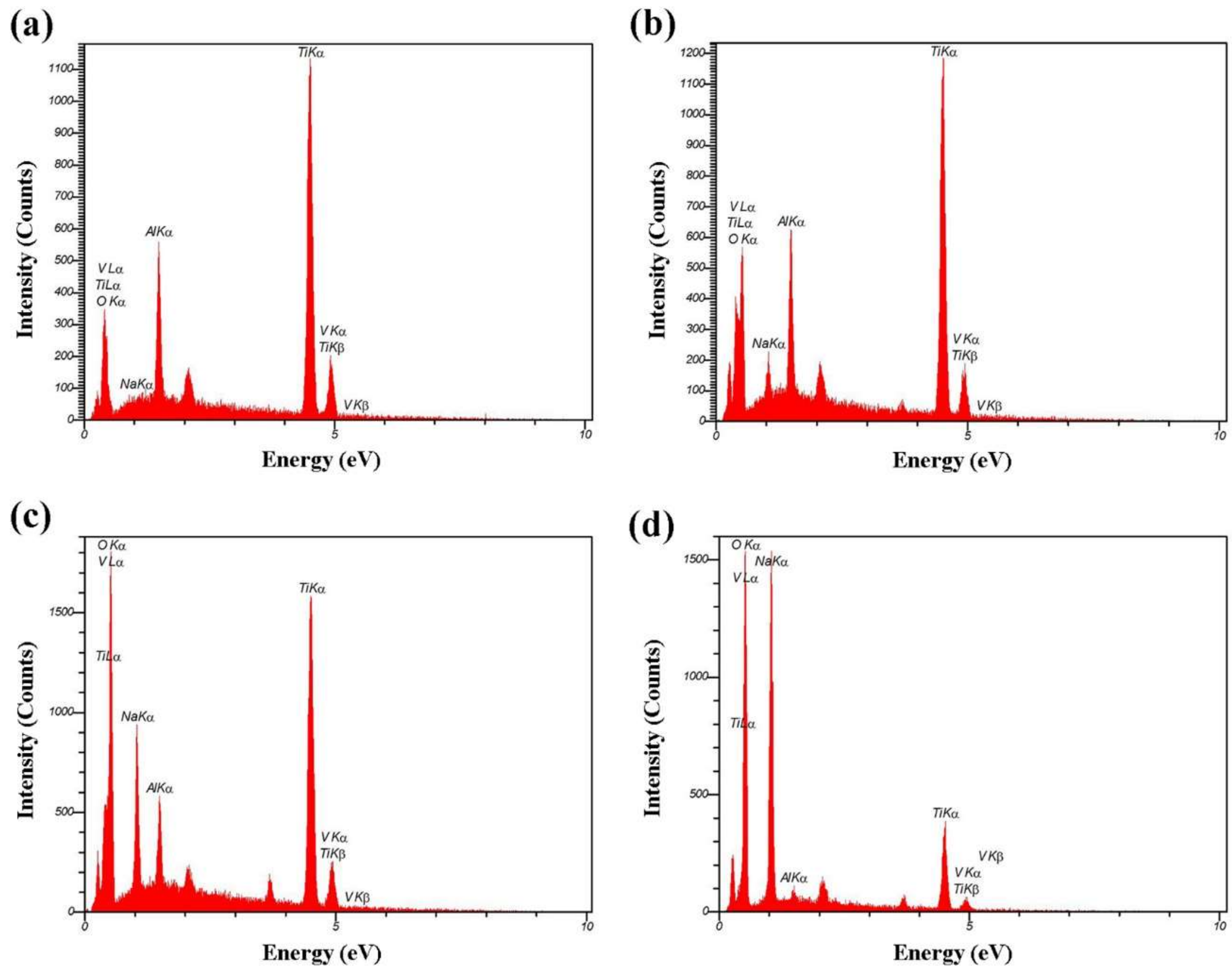

Figure 3. EDS profiles of Samples 1 (a), 3 (b), 5 (c) and 7 (d).

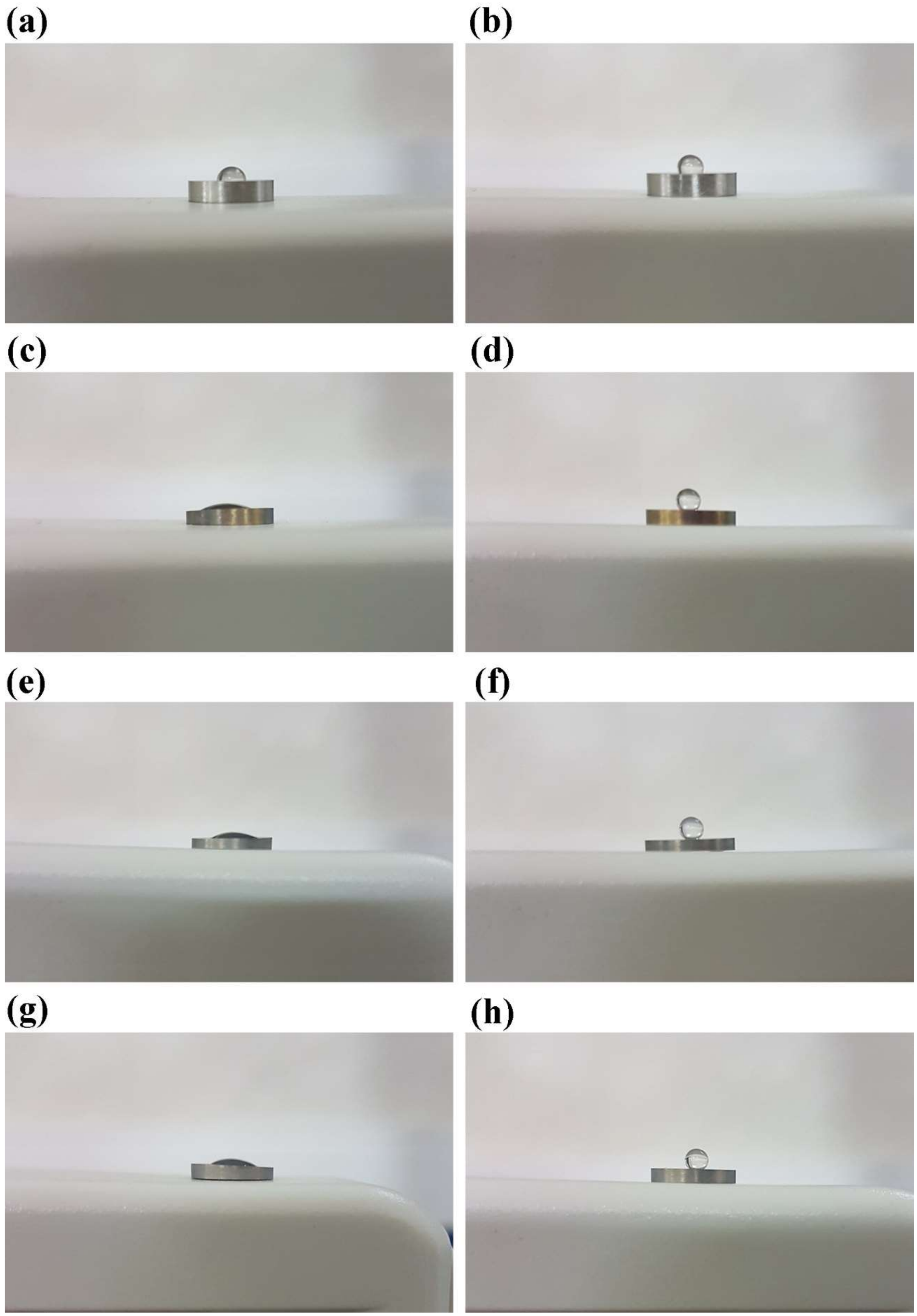
(a)
(b)
(c)
(d)
(e)
(f)
(g)
(h)

Figure 4. Macrographs of a water droplet on the surfaces of Samples 1 (a), 2 (b), 3 (c), 4 (d), 5 (e), 6 (f), 7 (g) and 8 (h).

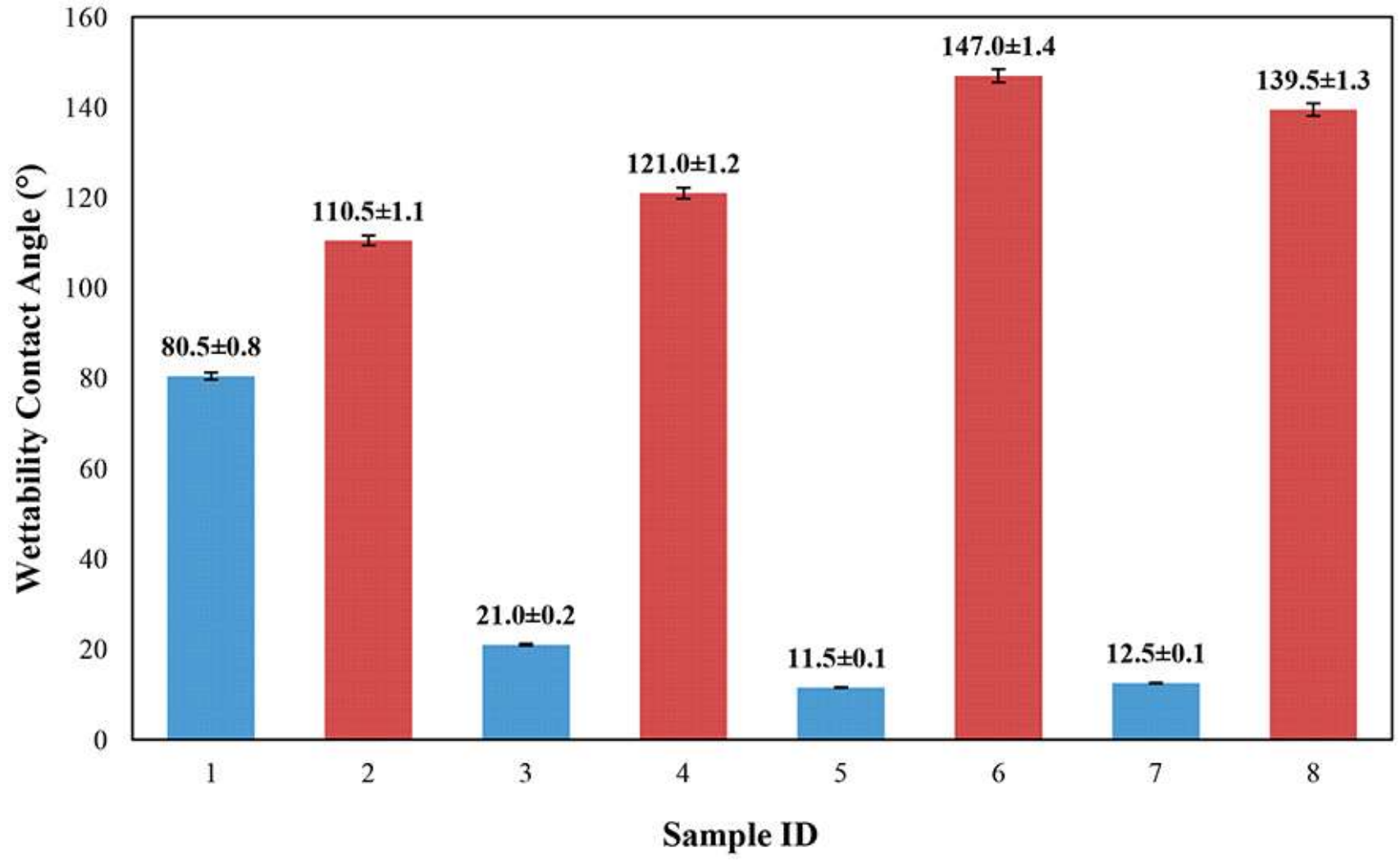

Figure 5. Amounts of the sessile water contact angles on the samples, in terms of "mean ± standard deviation".

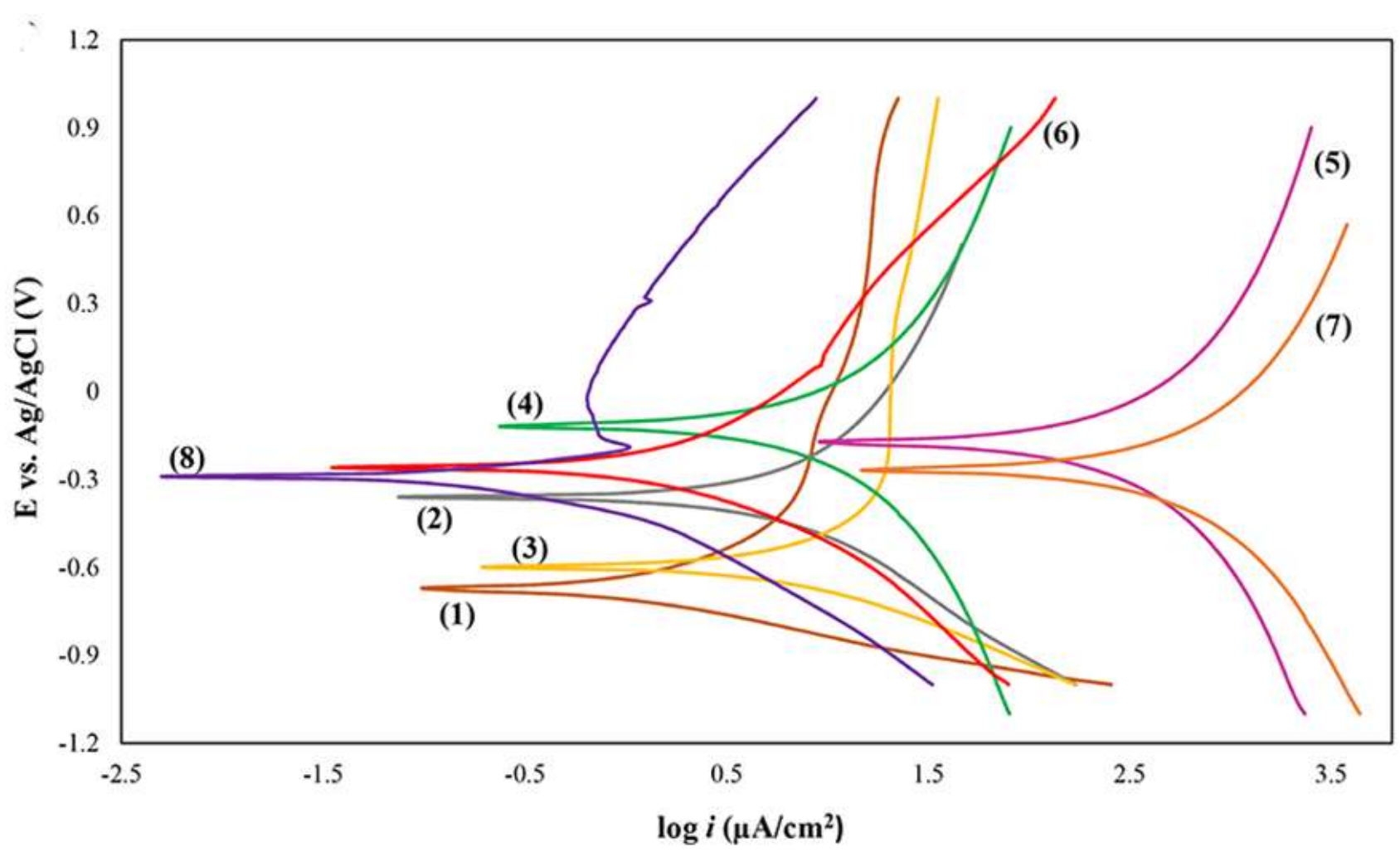

Figure 6. Potentiodynamic polarization curves of the samples.

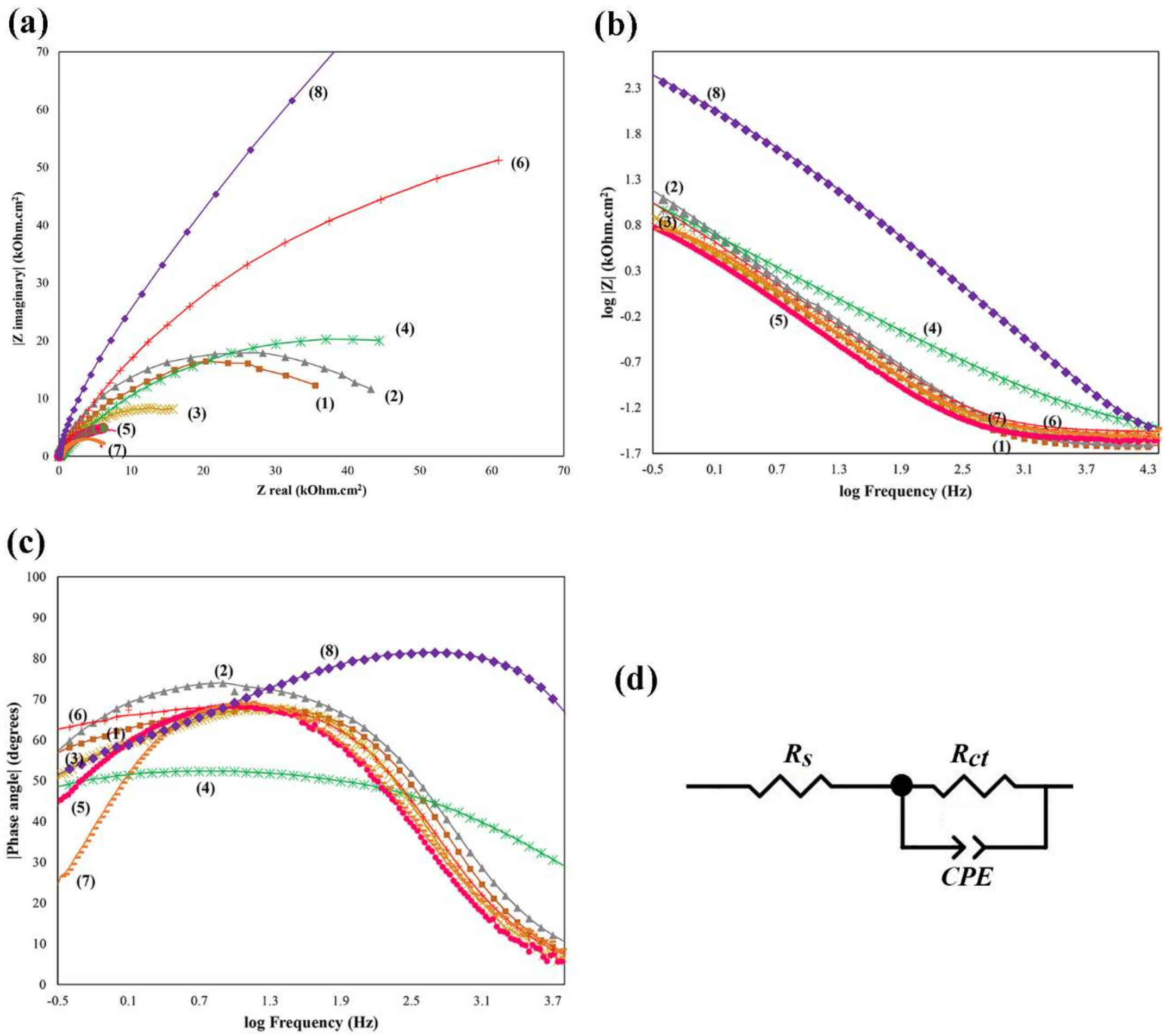

Figure 7. EIS Nyquist (a), Bode impedance (b) and Bode phase angle (c) plots of the samples and schematic of the used equivalent electrical circuit (d).

## Tables

**Table 1.** Samples coding as a function of the NaOH concentration of the alkaline treatment solution and the absence and presence of the HDTMS coating.

| **Specification** | **Sample ID** | | | | | | | |
|---|---|---|---|---|---|---|---|---|
| | **1** | **2** | **3** | **4** | **5** | **6** | **7** | **8** |
| Molar concentration of NaOH | - | - | 1 | 1 | 3 | 3 | 5 | 5 |
| HDTMS Coating | - | ✓ | - | ✓ | - | ✓ | - | ✓ |

**Table 2.** Results of the EDS analysis on the untreated and alkaline-treated surfaces.

| **Sample ID** | **Detected element (wt.%)** | | | | |
|---|---|---|---|---|---|
| | Ti | Al | V | Na | O |
| **1** | 89.7 | 7.2 | 1.6 | 0.0 | 1.5 |
| **3** | 82.4 | 6.1 | 1.0 | 2.1 | 8.4 |
| **5** | 54.1 | 3.7 | 0.6 | 5.2 | 36.4 |
| **7** | 27.6 | 1.8 | 0.2 | 18.0 | 52.4 |

**Table 3.** Corrosion potential ($E_{corr}$), corrosion current density ($i_{corr}$) and protection efficiency ($\eta$) extracted from the polarization diagrams. Nota that only the corrosion protection efficiency of the hydrophobized surfaces is reported, because the alkaline-treated samples present lower corrosion resistances compared to the untreated surface used as the reference.

| Sample ID | $E_{corr}$ (V) | $i_{corr.}$ (A/cm$^2$) | $\eta$ (%) |
|---|---|---|---|
| **1** | -0.6953 | 3.901E-5 | - |
| **2** | -0.3737 | 1.986E-6 | 94.9 |
| **3** | -0.6298 | 9.628E-5 | - |
| **4** | -0.1239 | 2.190E-7 | 99.4 |
| **5** | -0.1859 | 1.289E-4 | - |
| **6** | -0.2740 | 1.934E-7 | 99.5 |
| **7** | -0.2633 | 7.704E-4 | - |
| **8** | -0.2881 | 9.028E-8 | 99.8 |

**Table 4.** Fitting parameters of the equivalent electrical circuit modeling of the EIS data.

| **Sample ID** | $R_{ct}$ **(kΩcm$^2$)** | $Y_0$ **(nΩ$^{-1}$cm$^{-2}$ )** | $n$ | $C$ **(nFcm$^{-2}$)** | **Residual error (%)** |
|---|---|---|---|---|---|
| **1** | 50.48 | 51.16 | 0.82 | 63.01 | 2.35 |
| **2** | 56.77 | 49.83 | 0.83 | 61.67 | 3.66 |
| **3** | 26.32 | 78.91 | 0.78 | 96.98 | 3.34 |
| **4** | 81.78 | 44.56 | 0.85 | 55.98 | 3.84 |
| **5** | 14.67 | 91.45 | 0.77 | 99.84 | 5.27 |
| **6** | 152.05 | 35.43 | 0.86 | 46.60 | 2.64 |
| **7** | 7.11 | 113.09 | 0.75 | 105.16 | 9.12 |
| **8** | 533.45 | 2.33 | 0.91 | 2.38 | 2.80 |